\documentclass[twocolumn,prb,color,psfig,superscriptaddress]{revtex4}
\usepackage{graphicx}
\usepackage{epsfig}

\def\beq{\begin{equation}}
\def\eeq{\end{equation}}
\def\123{YBa$_2$Cu$_3$O$_{6+x}$}
\def\214{La$_{2-x}$Sr$_x$CuO$_{4}$}
\begin{document}

\title{
Nonbonding oxygen holes and spinless scenario of magnetic response in doped
cuprates }
\author{A.S. Moskvin}
\affiliation{Department of Theoretical Physics, Ural State University, 620083
Ekaterinburg, Russia}
\date{\today}

\begin{abstract}
Both theoretical considerations  and experimental data point to a more
complicated nature of the valence hole states in doped cuprates than it is
predicted by  Zhang-Rice model. Actually, we deal with a competition of
conventional hybrid Cu 3d-O 2p $b_{1g}\propto d_{x^2 -y^2}$ state and purely
oxygen nonbonding state with $e_{u}x,y \propto p_{x,y}$ symmetry. The latter
reveals a non-quenched Ising-like orbital moment that gives rise to a novel
spinless  purely oxygen scenario of the magnetic response in doped cuprates with the oxygen localized orbital magnetic moments  of the order of tenths of Bohr magneton.
We consider the mechanism of ${}^{63,65}$Cu-O 2p transferred orbital hyperfine
interactions due to the  mixing of the oxygen O 2p orbitals with Cu 3p semicore orbitals. Quantitative estimates point to a large
magnitude of the respective contributions both to local field and electric
field gradient, and their correlated character.
\end{abstract}

\maketitle

\section{Introduction}
The role played by magnetism, particularly the nature of magnetic fluctuations,
is one of the central issues of the high-T$_c$ cuprate physics.
Nuclear magnetic resonance (NMR) and nuclear quadrupole resonance (NQR) are
believed to provide the basic experimental grounds for a spin-fluctuation
mechanism of high temperature superconductivity. It is worth to note that namely the spin-lattice
relaxation rates and the Knight shift measurements by the NMR and NQR
stimulated the elaboration of the well known antiferromagnetic spin-fluctuation
scenario for the cuprates.\cite{Pines}
NMR first revealed a suppression of the low-energy magnetic excitations
below what is called the spin gap temperature. In the underdoped region, it is thought
that above T$_c$ a pseudo-gap opens up in the spin fluctuation spectrum. Since the
spin-gap state is believed to be related to the pairing mechanism, a large number of
experimental and theoretical studies have focused on the origin of the spin gap.
Despite some criticism \cite{PWA1} the spin-fluctuation scenario continues to
be very popular both in the NMR/NQR and HTSC community.
  However, a conventional approach to the hyperfine coupling and the nuclear resonance in cuprates implies a treatment within simple models usually applied to the conventional metals or vice versa to the weakly covalent and weakly
   correlated insulator.  The  magnetic response is assumed to be provided by the only contribution of the spin
degrees of freedom.  As in parent  antiferromagnetic oxides the Cu$^{2+}$ center with s=1/2 is considered to be a main
 resonating center whereas the doped holes are considered to form  an  usual Fermi-liquid.
  Meanwhile, a hole doping in the framework of the strongly-correlated scenario results in
   a formation of the well isolated Zhang-Rice ${}^{1}A_{1g}$ singlets.  The hyperfine interactions and NMR-NQR experiments in cuprates right up to now are interpreted within the Shastri-Mila-Rice (SMR) spin Hamiltonian.\cite{SMR}  A conventional approach to the analysis of the ${}^{63,65}$Cu NQR/NMR
experiments in the hole-doped cuprates corresponds to the model of uniform
lattice and indirectly implies  the 100$\%$ volume fraction of the equivalent
resonating nuclei.

Despite a great many of experimental and theoretical papers the nature and proper
description of the magnetic correlations in cuprates is still a subject of controversy.
 Results of the recent
NQR/NMR experiments  for "classic" cuprate systems 214 and 123 together with a
number of early data cast doubt on a validity of the popular concepts to be
widely used  as a starting point for analysis of the nuclear resonance and in a
more broad sense for many other physical effects. First,  it should be
noticed that the ${}^{63,65}$Cu NQR lines in the doped cuprates are
sometimes unusually inhomogeneously broadened ($2\div 4$ MHz), practically irrespective of the
doping level.\cite{Tsuda,Fujiyama,Song,Imai,Hammel,Goto,Martin,Itoh,Gippius}
 Experimental Cu NQR study in
\214, La$_2$CuO$_{4+\delta}$  has revealed two distinct
Cu(2) sites (A and B) with distinguishing relaxation rates and universal
difference in corresponding quadrupole frequencies.  Subsequently, a
precise measurement of the nuclear relaxation in \214  has
revealed a composite structure of the separate Cu NQR lines with
strong frequency dependence of  T$_1^{-1}$  across the spectrum.  At last, first Cu NQR measurements have revealed either an unexpectedly small value of the asymmetry parameter $\eta$  or rather
large difference of $\eta$ for A and B components. 
J. Haase et al. \cite{Haase1} have shown that the broadening of the Cu line in 214 system cannot be explained by spin effects and evidences  the orbital shift modulation of a short-length scale. The full planar oxygen spectra show a correlated modulation of the electric field gradient with the spin susceptibility.  NMR spin-echo double-resonance experiments uncovered the large distribution of the local magnetic fields at the planar Cu sites.\cite{Haase} 
 They found that a single fluid spin-only picture could not reproduce the experimental data.

Above we address mainly the NMR-NQR studies, however, a close inspection of
other magnetic data evidences the same controversies. The absence of an ESR
signal is strong evidence that local moments in cuprates do not exist. The
polarised neutron results presented by Smith {\it et al.} \cite{Smith} have
demonstrated that there is neither an elastic nor a quasi-elastic magnetic
response in the normal state of nearly optimally doped YBa$_2$Cu$_3$O$_{6.95}$.
Their data are inconsistent with the  existence of local spin magnetic
moments in the CuO$_2$ planes. Little scattering they observed can be assigned
to $\sim 3 \%$ of the Cu atoms carrying a spin $1/2$. They note that neither
the variation in magnitude of the susceptibility in 123 system with oxygen
content nor the temperature variation is consistent with the existence of local
moments.
 The integral intensity of the famous resonanse peak in 123 cuprate does not exceed 1-2 \% from that for 
spin-wave resonanse in parent system.\cite{Bourges} A drastic decreasing of the AF susceptibility amplitude  as a function of doping is found by INS, that disagrees with NMR data and questions the role of spin fluctuations in HTSC as the magnetic fluctuations seem to vanish for samples with largest T$_c$.\cite{Bourges}

Both the NMR-NQR and neutron measurements cannot discriminate between the spin
and orbital origin of electron magnetic moments. Thus, we cannot definitely
state that current experimental data unambiguously confirm the spin nature of
the magnetism in the doped cuprates. 
Furthermore, recently there appeared many indications to the orbital magnetism in cuprates. Possible formation
of antiferromagnetism below the
superconducting transition temperature was found
by several experimental techniques in underdoped \123 and
\214.\cite{Mook1,Mook2,sidis,sonier,lake} The relatively small
values of the observed magnetic moments \cite{Mook1,Mook2,sidis}
($0.01\div 0.05\beta _e$) have indicated
an orbital rather than a spin origin of the observed
antiferromagnetism. 
Most recent ARPES observation of the circular dichroism in the normal state of underdoped and overdoped Pb-Bi2212 samples \cite{Borisenko} also may be related to the persistent orbital currents. 

The NQR study provides a
more direct prove
for the formation of orbital magnetism since it is performed in
zero magnetic field. Thus, the internal magnetic moments if
they are present will result in an NQR line splitting. The first  experimental evidence for the formation of the internal magnetic
moments in the underdoped three-CuO$_2$-layer
Hg$_{0.8}$Cu$_{0.2}$Ba$_2$Ca$_2$Cu$_3$O$_{8+\delta}$ (Hg-1223)
high-T$_c$ cuprate superconductor  below $T_c$ = 134~K has been presented by Breitzke {\it et al.}\cite{Breitzke}
Using NQR technique they show that Cu NQR-lines split below
$T_c$ due to a Zeeman splitting
originating from the internal magnetic fields within the CuO$_2$-layers. These
results strongly favor a formation
of staggered orbital currents as an
origin of the observed phenomenon. The
values of the internal magnetic fields vary from the inner to the outer
CuO$_2$-layer and are of order of several hundred Gauss. Note, the fields
occur below $T_c$ and their intensities increase with decreasing
temperature.
The $^{199}$Hg Knight shift measurements in HgBa$_2$CuO$_{4+\delta}$ \cite{Suh}
have revealed very large anisotropic shifts which were assigned to orbital
magnetic moments $\mu \approx 0.04 \beta _e$ localized on the oxygen positions.
The ${}^{63,65}Cu$ shift distribution in La$_{1.85}$Sr$_{0.15}$CuO$_4$ is found recently to be of $orbital$ (!) origin. \cite{Haase}

 In our opinion,  these and many other experimental observations point to an inconsistency of a
conventional model of the well isolated spin and orbital Zhang-Rice (ZR)
singlet ${}^{1}A_{1g}$ \cite{ZR} believed to be a ground state of the
hole-doped CuO$_4$ center in the CuO$_2$ layers. Here, it should be noted that
when speaking of a Zhang-Rice singlet as being "well isolated", one implies
that the ${}^{1}A_{1g}$ ground state for the CuO$_4$ plaquette with the two
holes of the $b_{1g}(d_{x^2-y^2})$ symmetry is well separated from any other
excited two-hole states.  Both, experimental data and theoretical model
considerations evidence in favor of the more complicated structure of the
valence multiplet for the hole-doped CuO$_4$ center rather than simple ZR
singlet albeit namely the latter  is a guideline in the overwhelming majority
of current model approaches.

So, Y. Yoshinari {\it et al.}
\cite{Fisk} have undertook the Cu NQR study of the isolated hole
centers in La$_{2}$Cu$_{0.5}$Li$_{0.5}$O$_{4}$. Their results could be interpreted as
convincing evidence of the singlet-triplet structure of the hole center.  The
authors have revealed the spin singlet ground state (S=0) and the low lying
spin triplet state (S=1) with the singlet-triplet separation $\Delta _{ST} =
0.13$ eV which is comparable with the Cu-Cu nearest neighbor exchange integral
in parent oxide La$_2$CuO$_4$. 
An experimental
indication to the appreciable role of the O 2p$\pi$ orbitals in the ${}^{17}$O
hyperfine coupling was  obtained by Y.Yoshinari.\cite{Yo} This implies
a complicated nature of the ground state manifold for the CuO$_4$ center with a
significant mixing of the Zhang-Rice singlet and some other molecular term,
whose symmetry should be distinct from  ${}^{1}A_{1g}$. This conclusion
conflicts with the widespread opinion regarding the well isolation of the
Zhang-Rice singlet.

The nature of the valent hole states in doped cuprates is considered as being
of great importance for the high-$T_c$ problem. Having solved the problem we
could justify the choice of the relevant effective Hamiltonian together with
the opportunities of a mapping to the single band $t-J$ or Hubbard model.

Below we show that the outgoing beyond ZR model does predict a novel spinless
scenario of magnetic response in cuprates.

\section{A-E orbital structure of  hole CuO$_4$ centers in cuprates}

\begin{figure}[h]
\includegraphics[width=8.5cm,angle=0]{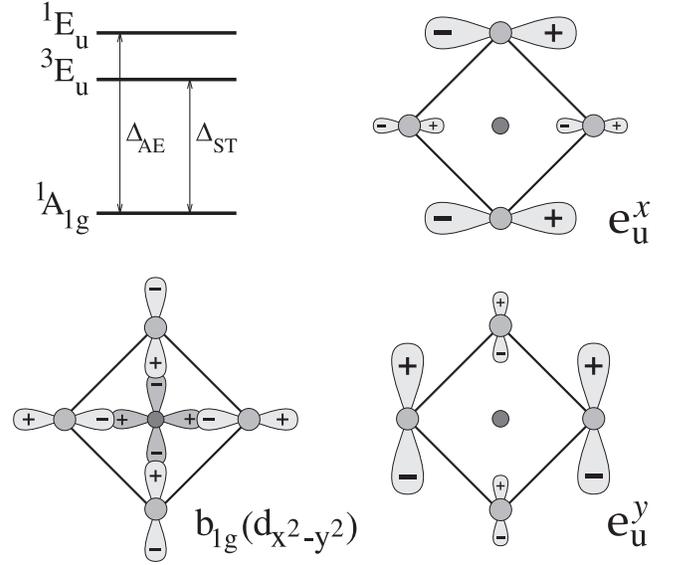}
\caption{The term  structure of the actual  valent A-E multiplet for hole
CuO$_{4}^{5-}$ center together with single-hole basis $b_{1g}^{b}$ and
$e_{ux,y}^{b}$  orbitals} \label{fig3}
\end{figure}

 Intrinsic nature of electron and hole centers in oxides is related to  the self-trapped charge transfer
 (CT) excitons.  Both
experimental observations and theoretical analysis point to a complex
two-component structure of the low-energy CT band near $2$ eV  in parent insulating cuprates. \cite{MoskvinPRB,MoskvinPRL} Here, we deal with a superposition of rather
well-defined one- and
 two-center CT excitons. The former is associated with a dipole-allowed
transition $b_{1g}\rightarrow e_u$ from the ground state $b_{1g}^{b}$ to the
purely oxygen non-bonding  doublet  $e_{u}(\pi)$ in the CuO$_4$ plaquette,
which is allowed in the "in-plane" polarization ${\bf E}\perp C_{4}$. The
latter is attributed to a $b_{1g}\rightarrow b_{1g}$ CT between two neighboring
CuO$_4$ plaquettes with the formation of electron CuO$_{4}^{7-}$ and hole
CuO$_{4}^{5-}$ centers. Here, the electron center nominally represents the
system of Cu$^{1+}$ and O$^{2-}$ ions with completely filled shells, whereas
the hole one does the system with two $b_{1g}^{b}$ holes forming the Zhang-Rice
(ZR) singlet.\cite{ZR} The one-center CT exciton formally consists of  the
conventional electron center and unconventional hole center with actually
two-hole configuration $b_{1g}e_u$ resulting in a spin singlet ${}^{1}E_u$ or
triplet term ${}^{3}E_u$, respectively. Both CT excitons can interact with each
other. Hence, to describe the $el-h$-structure of both excitons on an equal
footing one needs to consider the conventional electron center CuO$_{4}^{7-}$
and unconventional CuO$_{4}^{5-}$ hole center with actual $^1A_{1g},^{1,3}E_u$
multiplet. Hence, unlike a simple Zhang-Rice  model our model assumes a
quasi-degeneracy in the ground state of  hole CuO$_{4}^{5-}$ center with two
close in energy $^{1}A_{1g}$ (ZR-singlet) and ${}^{1,3}E_{u}$ terms of
$b_{1g}^2$ and $b_{1g}e_u$ configurations, respectively. This implies two near
equivalent locations for the additional hole, either in the Cu 3d-O 2p hybrid
$b_{1g}(d_{x^2 -y^2})$ state to form  ZR singlet ${}^{1}A_{1g}$, or in a purely
oxygen nonbonding doublet $e_{ux,y}$ state to form the ${}^{1,3}E_{u}$ state.
Fig.\ref{fig3} shows the term  structure of the actual {\it valent A-E
multiplet} for hole CuO$_{4}^{5-}$ center together with the single-hole basis
orbitals. These orbitals are defined as follows:
\begin{equation}
|b_{1g}^{b}\rangle =\cos \alpha _{b_{1g}} |b_{1g}(3d)\rangle + \sin
\alpha _{b_{1g}}|b_{1g} (2p)\rangle ,
\end{equation}
where $b_{1g}(3d)=3d_{x^2-y^2}$ and $b_{1g} (2p)$ are copper and oxygen molecular orbitals with $b_{1g}$ symmetry.
There are two types of purely oxygen nonbonding orbitals with $e_u$ symmetry: $e_{u}(\sigma)$ and $e_{u}(\pi)$, respectively, that  hybridize with each other (equally for both types ($x,y$) of such orbitals):
$$
|e_{u}^{b}\rangle =\cos \alpha _{e}\,|e_{u}(\pi)\rangle + \sin \alpha
_{e}\,|e_{u}(\sigma)\rangle ;
$$
\begin{equation}
|e_{u}^{a}\rangle =\sin \alpha _{e}\,|e_{u}(\pi)\rangle - \cos \alpha
_{e}\,|e_{u}(\sigma)\rangle \, ,
\label{eu}
\end{equation}
where
\begin{equation}
\tan 2\alpha _{e} =\frac{2t^{pp}_{e_{u}}}{\epsilon _{pe_{u}(\sigma)}- \epsilon
_{pe_{u}(\pi)}}\, , \label{tan}
\end{equation}
and
$$
t^{pp}_{e_{u}}=-(t_{pp\sigma}+t_{pp\pi})
$$
is an effective transfer integral with $t_{pp\sigma}<0$, $t_{pp\pi}>0$ being
 two types of $pp$ transfer integrals, for
$\sigma$ and  $\pi$ bonding, respectively ($|t_{pp\pi}|\approx \frac{1}{2}
|t_{pp\sigma}|$). Hereafter, we preserve the notations
$e_{u}(\sigma),e_{u}(\pi)$  for dominantly $\sigma$, or $\pi$ orbital,
respectively.
 Interestingly, that
 $e_{u}(\sigma),e_{u}(\pi)$ orbitals could form two types of
 circular current $p_{\pm 1}$-like states: $e_{u\pm 1}(\sigma),e_{u\pm 1}(\pi)$,
  respectively, with Ising-like orbital moment
\begin{equation}
\langle e_{u\pm 1}(\pi)|l_z|e_{u\pm 1}(\pi)\rangle = -\langle e_{u\pm
1}(\sigma)|l_z|e_{u\pm 1}(\sigma)\rangle =\pm \sin2\alpha _e 
\end{equation}
 or two types of currentless $p_{x,y}$-like $e_{ux,y}(\sigma),e_{ux,y}(\pi)$ states
 with a quenched orbital moment.
The A-E model with a $b_{1g}-e_u$ competition goes essentially beyond the
well-known ZR model. In a sense, the valence
$(b_{1g}^{2}){}^{1}A_{1g}-(b_{1g}e_{u}){}^{1,3}E_{u}$ multiplet for the hole
 center represents an unconventional state with Cu valence
resonating between Cu$^{3+}$ and Cu$^{2+}$, or "ionic-covalent" bonding.
 In other words, we deal with  a specific version of the "correlation" polaron, introduced by
Goodenough and Zhou. \cite{Goodenough} Such centers are characterized by strong
coupling with lattice and can reveal the (pseudo)Jahn-Teller effect.\cite{2}

The orbital doublet terms ${}^{1,3}E_{u}$ for  hole CuO$_{4}^{5-}$ center are
straightforwardly derived from two-hole $b_{1g}e_{u}$ configuration, whereas
the configurational interaction is surely to be taken into account when
deriving the ZR singlet ${}^{1}A_{1g}$. For the reasonable values of parameters
(in eV): $U_{d}= 8.5$, $U_{p}=4.0$, $V_{pd}=1.2$, $\epsilon _{d}=0$, $\epsilon
_{p}=3.0$, $t=1.3$ \cite{MoskvinPRB} its wave function can be written as follows
$$
\Psi _{1}=|ZR\rangle = -0.25|d^{2}\rangle +0.95|dp\rangle  -0.19|p^{2}\rangle ,
$$
where three $b_{1g}^{2}$-like  configuration are mixed. This function reflects
the well-known result that the ZR-singlet represents a two-hole configuration
with one predominantly Cu 3d and one predominantly O 2p holes, however, having the same $b_{1g}$ symmetry.

The $b_{1g}-e_u$ hole competition reflects the subtle balance between the gain in electron-electron repulsion (U$_{dd}>$V$_{pd}$) and  the loss in one-particle energy both affected by a lattice polarization. The $b_{1g}-e_u$, or A-E model is  supported both by
local-density-functional calculations \cite{McMahan1} and {\it ab initio}
unrestricted Hartree-Fock self-consistent field MO method (UHF-SCF) for
copper-oxygen clusters.\cite{Tanaka,Tanaka1} To the best of our knowledge the one of
the first quantitative conclusions on a competitive role of the hybrid
copper-oxygen $b_{1g}(d_{x^2 -y^2})$ orbital and purely oxygen O 2p$_{\pi}$
orbitals in the formation of valent states near the Fermi level in the CuO$_2$
planes has been made by A.K. McMahan {\it et al.}  \cite{McMahan1} and J.
Tanaka {\it et al.}\cite{Tanaka,Tanaka1}  Namely these orbitals, as they state,
define the low-energy physics of copper oxides.

In connection with the  valent $^{1}A_{1g}-{}^{1,3}E_{u}$ manifold model for
copper oxides one should note and comment the results of paper by Tjeng {\t et
al}. \cite{Tjeng} where the authors state that they "...are able to unravel the
different spin states in the single-particle excitation spectrum of
antiferromagnetic  CuO and show that the top of the valence band is of pure
singlet character, which provides strong support for the existence and
stability of Zhang-Rice singlets in high-$T_c$ cuprates". However, in their
photoemission work they made use of the Cu 2p$_{3/2}(L_{3})$ resonance condition
that allows to detect unambiguously only copper photohole states, hence they
cannot see the purely oxygen photohole $e_u$ states.

Interestingly to note, that among three possible states for trapped hole in
cuprate: ZR singlet ${}^{1}A_{1g}$, spin singlet ${}^{1}E_{u}$, and spin
triplet ${}^{3}E_{u}$, only the latter provides relevant conditions for the
hole transport through antiferromagnetic background. In other words, one might
speak about the spin-triplet channel of $e_{u}(\pi)$ hole transport
 as a main mechanism of conductivity in insulating cuprates.\cite{MR}

 \subsection{Unconventional magneto-electric  CuO$_4$ hole centers 
beyond simple ZR singlet picture.}

Unconventional orbital A-E structure of the hole CuO$_{4}^{5-}$ center in EH
droplet goes beyond simple ZR singlet picture and deserves more close
examination. 
Neglecting the spin degree of freedom we  introduce a pseudo-spin formalism to
describe the  orbital states of
 the CuO$_4$ centers in the framework of the  valent ($^{1}A_{1g},{}^{1}E_{u}$) multiplet model.
   Three orbital states of the ($^{1}A_{1g},{}^{1}E_{u}$) multiplet we associate
 with three states of {\it orbital pseudo-spin} $S =1$:
$
|z\rangle =|^{1}A_{1g}\rangle;|x,y\rangle =|^{1}E_{u}x,y\rangle.
$
Then the pseudospin matrix  has   a very simple form:
$
\langle i |\hat S_{k} | j \rangle =i \epsilon _{ikj}.
$
A complete set of the pseudo-spin operators should include both
 ${\bf  S}$ and five spin-quadrupole operators
 $$
  \widetilde{\{{\hat S_{i}},{\hat S_{j}}\}}=\{{\hat S_{i}},{\hat S_{j}}\}-\frac{2}{3} {\hat
{\bf S}}^{2}\delta _{ij}\, .
$$
These pseudo-spin operators are not to be confused with real physical
spin-operators as they act in a pseudo-space. Nevertheless, all these  correspond to real
physical quantities. First,
 the $z$-component of
pseudo-spin defines the only non-zero $z$-component of the Ising-like orbital
magnetic moment:
 ${\hat {\bf M}} = \hat{g}^{M}{\hat {\bf S}}$,  with the only nonzero $g_{zz}$
component of  $g^{M}$ -tensor.
Microscopically, the effective magnetic moment is generated by the orbital
currents for the $e_u$ hole. Taking into account only local oxygen
contributions one may write
$$
{\hat {\bf M}}= \beta _{e}\sum _{n=1}^{4} {\hat {\bf l}}_{n}\, ,
$$
and
$$
g_{zz}^M = \imath \beta _{e} \langle E_{u}x|\sum _{n=1}^{4} {\hat l}_{nz}|E_{u}y\rangle \, ,
$$
where ${\hat {\bf l}}_{n}$  is the  orbital momentum operator for $n$-oxygen.
 Second, the $S_{x,y}$ pseudo-spin components
define the unconventional quantity with spatio-transformational properties of
polar vector like electric field, and time-inversion symmetry like magnetic
field. This is a so-called toroidal moment which can be defined for the CuO$_4$
plaquette as follows:
$
 {\hat {\bf T}} = [{\bf g}^{T}\times{\hat {\bf S}}]$,
 where the $g^{T}$-vector has the  only non-zero $z$-component.
Microscopically, the effective toroidal moment can be derived through the
local oxygen effective orbital moments as follows:
$$
{\hat {\bf T}}= \beta _{e}\sum _{n=1}^{4} [{\bf R}_{n}\times{\hat {\bf l}}_{n}]\,,
$$
and
\begin{equation}
g_{z}^T = \imath \beta _{e} \langle A_{1g}|\sum _{n=1}^{4} R_{nx}{\hat
l}_{nz}|E_{u}y \rangle \, . \label{T}
\end{equation}
 It should be emphasized that both magnetic
 and toroidal moment are generated by the orbital currents for the oxygen
  holes. The numerical  magnitude of the effective orbital magnetic moment
   in $E_{u}$ state is determined mainly by the mixing of O 2p$\pi$ and
    O 2p$\sigma$ orbitals (see Exps. (\ref{eu}) and (\ref{tan})).
$$
 g^{M}_{zz} = \beta _{e}\sin 2\alpha _{e},
$$
where $\sin \alpha _{e}$  is a covalency parameter for
$e_{u}(\pi)-e_{u}(\sigma)$ bond.
 For a relatively small $\pi -\sigma$-mixing
$$
 g^{M}_{zz}\approx \beta _{e}\tan 2\alpha _{e} =\frac{2\beta _{e}t^{pp}_{e_{u}}}{\epsilon _{pe_{u}(\sigma)}-
\epsilon _{pe_{u}(\pi)}} \approx 0.2\beta _{e}
$$
given the reasonable values $t^{pp}_{e_{u}}\approx 0.3$ eV and $|\epsilon
_{pe_{u}(\sigma)}- \epsilon _{pe_{u}(\pi)}|\approx 3.0$ eV. For the
$g^{T}$-vector we readily obtain
$$
  g^{T}_{z} =\frac{1}{\sqrt{2}}
 \beta _{e}R_{CuO}\cos\alpha _{e}\sin \alpha _{b_{1g}},
$$
where $\sin \alpha _{b_{1g}}$ is a covalency parameter for
$b_{1g}(3d)-b_{1g}(2p)$ bond. This expression together with (\ref{T}) implies
that the toroidal moment is generated by oxygen orbital moments
$$
l_z = \frac{1}{2\sqrt{2}}\beta _{e}\cos\alpha _{e}\sin \alpha _{b_{1g}}\, ,
$$
which value can be estimated to be of the order of $0.2\beta _{e}$ given $|\sin
\alpha _{b_{1g}}|\approx 0.6$. It is quite probable that the toroidal
fluctuations will be  comparable, or even more pronounced than that of
conventional magnetic moment. The toroidal moment is distributed on CuO$_4$
plaquette and produces a nonzero  dipole magnetic field.
For all points lying in the CuO$_4$  plane the field  has $c$-axis orientation
whereas it has $ab$-orientation for all points lying in other symmetry planes.

Above we estimated the maximal values of magnetic and toroidal moments for the
A-E model of CuO$_4$ center. Puzzlingly, these compete with Cu$^{2+}$ spin magnetic moments in parent oxides, which are markedly reduced by a quantum reduction and covalent effects. Actually, we should deal with the quenching effect
of ``single-ion'' anisotropy or other crystalline fields on the orbital magnetism.

The symmetric quadratic pseudo-spin operators define  effective electric dipole
and quadrupole moments. The former has a planar character with  two non-zero
components: ${\hat d}_x = d_0\{\hat S_x\hat S_z\},{\hat d}_y = d_0\{\hat
S_y\hat S_z\}$,
 where $d_0$ is effective
dipole moment length. The latter has three non-zero components: $ \hat
Q_{A_1}=Q_{A_1}(\hat S_z^2-\frac{2}{3}),\, \hat Q_{B_1}=Q_{B_1}(\hat S_x^2-\hat
S_y^2),\, \hat Q_{B_2}=Q_{B_2}\{\hat S_x\hat S_y\}$
 with three quadrupole parameters $Q_{\Gamma}$. Thus,  the CuO$_4$
plaquette with ($^1A_{1g},^1E_u$) valent multiplet forms an unconventional
magneto-electric center characterized by eight independent orbital order
parameters. Generally speaking,  our model represents a theory that predicts
 broken  time-reversal ($T$) symmetry, two-dimensional parity ($P$),
 and basic tetragonal symmetry.

\subsection{Oxygen holes and orbital hyperfine interactions beyond the Shastry-Mila-Rice model}
Below we address some unconventional properties of ${}^{63,65}$Cu hyperfine
interactions for the spin-singlet $^{1}A_{1g}-{}^{1}E_{u}$  valence multiplet
of the CuO$_4$ center resulting from its non-quenched orbital moment.

The nuclear resonance experiments right up to now are interpreted within the
Shastri-Mila-Rice  spin-Hamiltonian  \cite{SMR}
\begin{equation}
{\hat H}_{hf} = \sum _{mn}{}^{63}{\bf I}(n)[{\hat A}(n){\bf s}(n)+B(nm){\bf
s}(m)],
\end{equation}
 based on the assumption that the spin density in the
CuO$_4$ centers is localized on the copper ions. Here ${\hat A}(n)$ is the
hyperfine tensor for the direct, on-site coupling of the ${}^{63,65}$Cu nuclei
to the Cu$^{2+}$ spins ($s=1/2$), $B(nm)$ is the strength of the transferred
hyperfine coupling of the ${}^{63,65}$Cu nuclear spin to the four nearest
neighbor Cu$^{2+}$ spins.

Effective Hamiltonian of nuclear quadrupole interactions for  $^{63,65}$Cu
nuclei has a conventional form as follows
\begin{widetext}
\begin{equation}
\hat H_Q = \frac{Q}{4I(2I-1)}\bigl [V_{zz}(3{\hat I}_{z}^{2}-{\hat {\bf
I}}^{2})+\eta V_{zz}({\hat I}_{x}^{2}-{\hat I}_{y}^{2})+\epsilon V_{zz}({\hat
I}_{x}{\hat I}_{y}+{\hat I}_{y}{\hat I}_{x})\bigr],
\end{equation}
\end{widetext}
where for CuO$_4$ center
$$
V_{zz}=V_{zz}({\bf R})=(V_{zz}^{E}-V_{zz}^{A}+V_{zz}^{p})\langle {\hat
S}_{z}^{2}\rangle _{{\bf R}} +V_{zz}^{A},
$$
$$
\eta V_{zz}=3V_{zz}^{p} \langle \hat S_{x}^{2}-\hat S_{y}^{2}\rangle _{{\bf
R}},\qquad \epsilon V_{zz}=3V_{zz}^{p} \langle \{\hat S_{x},\hat S_{y}\}\rangle
_{{\bf R}},
$$
where  ${\bf R}$ is the radius-vector of CuO$_4$ center. Parameters
$V_{zz}^{A}$ and  $V_{zz}^{E}$ determine the $b_{1g}$ contribution to $V_{zz}$
for $^{1}A_{1g}$ and $^{1}E_{u}$ terms, respectively, while $V_{zz}^{p}$ does
the total contribution of the Cu $p$ electrons. The $^{63,65}$Cu NQR frequency
can be written as follows
$$
\nu _{Q}=\frac{1}{2}|QV_{zz}|\sqrt{1+\frac{1}{3}(\eta ^{2}+\epsilon ^{2})}\, .
$$
 A variety of the model EFG calculations were carried out. \cite{Yu,Eremin,Husser,Kupcic}
  First,
we would like to note the extreme sensitivity of the EFG to the calculated
anisotropic charge distribution of the semicore Cu 3p states which are
characterized by the very large magnitude of the  effective quadrupole
parameter $\langle 1/r^3 \rangle _{3p}\simeq 150$ (Ref. \onlinecite{Yu}), or
$170\, a.u.$ (Ref. \onlinecite{Husser}). This parameter governs the magnitude
both of the EFG tensor and local magnetic field induced by Cu 3p electron on
copper nucleus:
\begin{equation}
V_{ij}=-\frac{2e}{5}\langle 1/r^3 \rangle _{3p}\langle (3\widetilde{{\hat
l}_i{\hat l}_j} - 2\delta _{ij}) \rangle ;
\end{equation}
\begin{equation}
{\bf H}_{loc}=-2\beta _{e}\langle 1/r^3 \rangle _{3p}{\bf l} \, ,
\end{equation}
where ${\bf l}$ is an orbital momentum for Cu 3p electron, and $\widetilde{{\hat l}_i{\hat l}_j}=1/2({\hat l}_i{\hat l}_j+{\hat
l}_j{\hat l}_i)$. Thus, the Cu 3p contribution to the EFG and to the local
field can reach colossal values such as 100 (in $10^{22}$ Vm$^{-2}$) and $10^3$
$Tesla$, respectively. In  conventional cuprate scenarios with valence
$b_{1g}\propto d_{x^2 -y^2}$ holes there is no hybridization between Cu 3p and
valence states,  and  the semicore Cu 3p contribution to electric and magnetic
hyperfine interactions can be taken into account in frames of Sternheimer
shielding-antishielding effects. However, the semicore Cu 3p role becomes
 of particular significance for
the $^{1}A_{1g},{}^{1}E_{u}$ valence multiplet of electron and hole centers
with varying hole density in oxygen $e_u$ states which have the same symmetry
as Cu 3p$_{x,y}$ states, that is these can hybridize with each other. As a
result, the purely oxygen $e_u$ orbital turns into O 2p-Cu 3p hybrid molecular
orbital (MO)
$$
\phi ^{e_u}_{x,y}\rightarrow \Phi ^{e_u}_{x,y}=c_{2p}\phi ^{e_u}_{x,y}+
c_{3p}\phi ^{3p}_{x,y}
$$
with MO coefficients $c_{3p}\ll c_{2p}$. Thus we arrive at the effective
magnetic and electric  "oxygen-to-copper" transferred orbital hyperfine
interaction. The effective $e_{u}(\pi)$ contribution to the local field on the
copper nucleus can be written in terms of pseudo-spin formalism as (in Tesla)
$$
 H^z_{loc}=-2\beta _{e}\langle 1/r^3 \rangle _{3p}|c_{3p}(\pi)|^2 \langle
S_z \rangle
$$
\begin{equation}
\approx 2.0\cdot10^3 |c_{3p}(\pi)|^2 \langle S_z \rangle \label{H}
\end{equation}
irrespective of
 the magnitude of the orbital moment for CuO$_4$ center. For the nonzero EFG components
$V_{zz},V_{xx},V_{yy},V_{xy}$ we obtain (in $10^{22}$ Vm$^{-2}$)
$$
V_{ij}=-\frac{2e}{5}\langle 1/r^3 \rangle _{3p}|c_{3p}(\pi)|^2\langle
(3\widetilde{{\hat S}_i{\hat S}_j} - 2\delta _{ij}) \rangle
$$
\begin{equation}
 \approx  2.7\cdot10^2 |c_{3p}(\pi)|^2\langle
(3\widetilde{{\hat S}_i{\hat S}_j} - 2\delta _{ij}) \rangle .
 \label{V}
\end{equation}
Interestingly, that Eqs.(\ref{H}) and (\ref{V}) imply that the ratio between
local field and EFG is governed only by the ratio between respective
pseudo-spin averages:
\begin{equation}
H^z_{loc}\,:\,V_{ij}= \beta _{e}\langle S_z \rangle \,:\,\frac{e}{5} \langle
(3\widetilde{{\hat S}_i{\hat S}_j} - 2\delta _{ij}) \rangle ).
\end{equation}
Simple relation between local field and EFG governed only by the respective
pseudo-spin averages implies a rather subtle interplay between magnetic and
electric contributions both to NMR-NQR frequencies and the spin-lattice
relaxation rate for copper nuclei.
 The numerical calculations  allow
us to expect the O 2p-Cu 3p mixing coefficient $c_{3p}$ to be of the order of
several hundredth. Indeed, the overlap contribution to this coefficient given
the Cu-O separations $R_{CuO}\approx 1.9 \AA $   is estimated  \cite{Eremin1} to
be $c_{3p}(overlap)=S_{Cu 3p-O 2p}^{\sigma}\approx -0.05$  for the strongest Cu
3p - O 2p $\sigma$-bonding and $S_{Cu 3p-O 2p}^{\pi}\approx
-0.5S_{Cu 3p-O 2p}^{\sigma}$. In such a way the oxygen $e_{u}(\pi)$ hole
contribution to   the orbital hyperfine interactions due to the
Cu 3p$\pi$-O 2p$\pi$ overlap can be estimated as $|H_{loc}|\leq 1$ Tesla and
$|V_{ij}|\leq 0.3\cdot
 10^{22}$ Vm$^{-2}$ for magnetic and electric terms, respectively. It should be noted
  that the respective maximal values correspond to the very large
 magnitude of effective NMR and NQR frequencies of the order of 10 MHz.
 Moreover, the oxygen $e_{u}(\sigma)$ hole contribution can be approximately four times
bigger.

\section{Conclusion}
We showed that outgoing beyond a simple ZR model we arrive at a complex $^{1}A_{1g}-{}^{1,3}E_{u}$ structure of the valent multiplet for the hole CuO$_4^{5-}$ center in cuprate with  engaging orbital degree of freedom. Moreover,
it should be emphasized that simple $^{1}A_{1g}-{}^{1}E_{u}$  model implies a spinless
purely orbital and purely oxygen  scenario of magnetic response and hyperfine
interactions in doped cuprates. However, we do not completely  reject the spin
degree of freedom. Indeed, our model implies a near degeneracy for singlet
${}^{1}E_{u}$ and triplet ${}^{3}E_{u}$ terms with many interesting
manifestations of the spin singlet-triplet magnetism.\cite{MoskvinSTM} Moreover, both
spin and orbital degrees of freedom are likely to be involved into a formation of
the complex magnetic response of doped cuprates with a relative weight that
manifests itself diversely depending on the energy range and experimental
conditions (NMR-NQR, magnetic susceptibility, magnetic neutron scattering,...).

 \section{Acknowledgments}
 The author acknowledges valuable discussions with S.V. Verkhovsky, M.V. Eremin, A.A. Gippius, A.V. Dooglav, J. Haase, and a partial support from the INTAS Grant No. 01-0654, CRDF Grant No. REC-005,  RME Grants No. E 02-3.4-392 and
No. UR.01.01.062, RFBR Grant No. 04-02-96077.

\end{document}